\documentclass[a4paper]{article}

\usepackage[utf8]{inputenc}
\usepackage[T1]{fontenc}
\usepackage{graphicx} 
\usepackage{wrapfig}
\usepackage{amsmath}
\usepackage{subcaption}
\usepackage{xcolor}
\usepackage{authblk}
\usepackage{hyperref}
\hypersetup{
  colorlinks=true,
  linkcolor=blue,
  citecolor=blue,
  urlcolor=blue
}

\title{\textbf{Dissecting Quantum Reinforcement Learning: A Systematic Evaluation of Key Components}}

\author{}

\date{}

\begin{document}

\maketitle

\vspace{-1.5cm}

\begin{center}
\begin{minipage}{0.28\linewidth}
\centering
\textbf{Javier Lazaro}\\[3pt]
\textit{University of Deusto}\\
\textit{Fsas International Quantum Center (Fujitsu)}\\
Bilbao, Spain\\
{\scriptsize \texttt{javier.lazaro@opendeusto.es}}
\end{minipage}
\hspace{0.01\linewidth}
\begin{minipage}{0.28\linewidth}
\centering
\textbf{Juan-Ignacio Vazquez}\\[3pt]
\textit{University of Deusto}\\
Bilbao, Spain
{\scriptsize \texttt{ivazquez@deusto.es}}
\hspace{0.01\linewidth}
\end{minipage}
\begin{minipage}{0.28\linewidth}
\centering
\textbf{Pablo Garcia Bringas}\\[3pt]
\textit{University of Deusto}\\
Bilbao, Spain\\
{\scriptsize \texttt{pablo.garcia.bringas@deusto.es}}
\end{minipage}
\end{center}

\vspace{0.4cm}

\begin{abstract} 
Parameterised quantum circuit (PQC) based Quantum Reinforcement Learning (QRL) has emerged as a promising paradigm at the intersection of quantum computing and reinforcement learning (RL). By design, PQCs create hybrid quantum–classical models, but their practical applicability remains uncertain due to training instabilities, barren plateaus (BPs), and the difficulty of isolating the contribution of individual pipeline components. In this work, we dissect PQC-based QRL architectures through a systematic experimental evaluation of three aspects recurrently identified as critical: (i) data-embedding strategies, with \emph{Data Reuploading} (DR) as an advanced approach; (ii) ansatz design, particularly the role of entanglement; and (iii) post-processing blocks after quantum measurement, with a focus on the underexplored \emph{Output Reuse} (OR) technique. Using a unified PPO–CartPole framework, we perform controlled comparisons between hybrid and classical agents under identical conditions. Our results show that OR, though purely classical, exhibits distinct behaviour in hybrid pipelines, that DR improves trainability and stability, and that stronger entanglement can degrade optimisation, offsetting classical gains. Together, these findings provide controlled empirical evidence of the interplay between quantum and classical contributions, and establish a reproducible framework for systematic benchmarking and component-wise analysis in QRL.
\end{abstract}


\section{Introduction}
\label{sec:intro}

\begin{figure}[t]
  \centering
  \includegraphics[width=\textwidth]{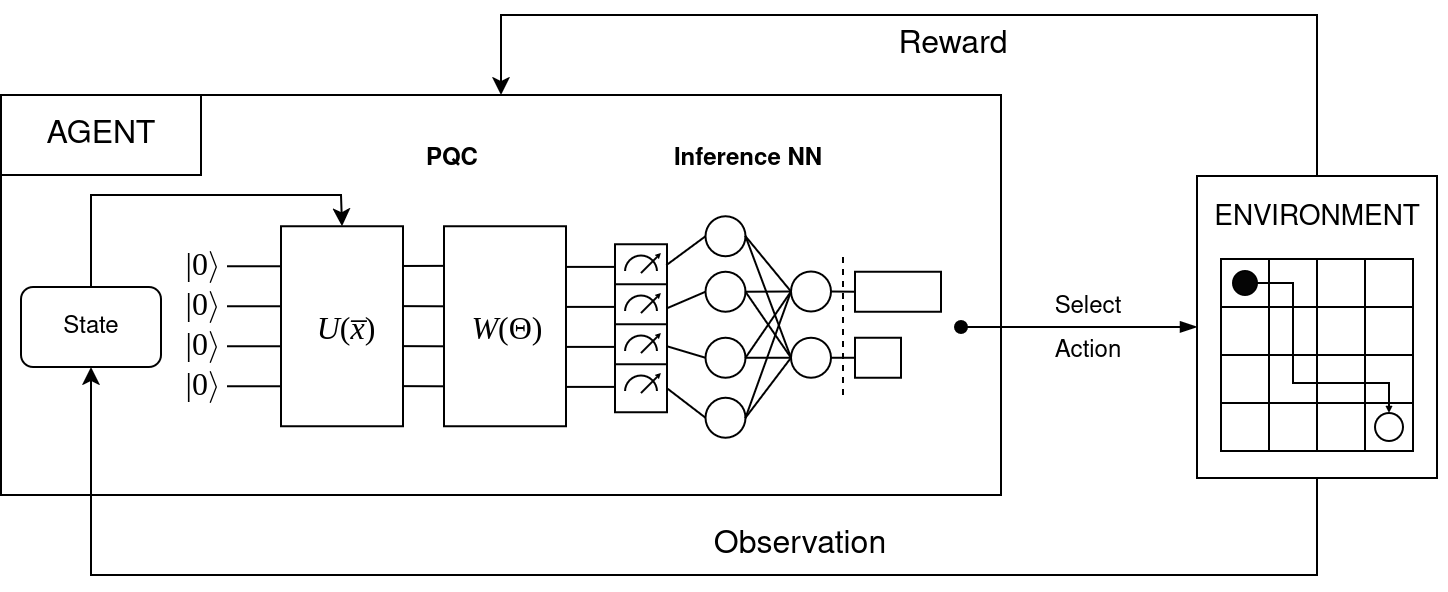}
  \caption{Schematic of a standard hybrid QRL pipeline integrating a parameterised quantum circuit (PQC) into a classical RL loop. The diagram highlights the three principal components analysed in this work: data embedding $U(\bar{x})$, variational ansatz $W(\Theta)$, and post-measurement inference, through which observations are encoded, processed quantum-mechanically, and interpreted classically to produce actions.}
  \label{fig:qrl-loop}
\end{figure}

Parameterised quantum circuit (PQC)-based Quantum Reinforcement Learning (QRL) represents a growing line of research aimed at integrating quantum resources into reinforcement learning algorithms. PQCs naturally give rise to hybrid quantum–classical models, where quantum circuits are embedded within a classical reinforcement learning pipeline. This hybrid setting has the potential to provide richer state representations and improved trainability, but it also introduces new sources of instability. Several recent works have assessed the practicality of QRL with PQCs, reporting both promising outcomes and significant limitations~\cite{skolik_robustness_2023,coelho_vqc-based_2024,hohenfeld_quantum_2024,kolle_study_2024,meyer_robustness_2024}. Across this literature, a few aspects consistently emerge as critical to performance: the role of the inference layer after measurement (often implemented through small neural networks as interpretation functions), the use of advanced embedding strategies such as Data Reuploading (DR), and the design of variational circuits, particularly ansatz architecture and entanglement~\cite{hsiao_unentangled_2022,perez-salinas_data_2020,kolle_study_2024}. Although these aspects have been repeatedly identified as bottlenecks, their individual contributions remain unclear, as they are rarely examined under consistent experimental settings. Fig.~\ref{fig:qrl-loop} sketches the hybrid pipeline analysed in this work and highlights the three blocks studied.

% --- FIGURA DE CONTEXTO DEL PIPELINE QRL ---

This lack of controlled analysis makes it difficult to determine which part of the pipeline is responsible for observed improvements or limitations, obscuring our understanding of QRL and hindering the development of principled design practices. To address this issue, we conduct a systematic evaluation of QRL models across the three components above under a unified and reproducible setting. All experiments are performed on the CartPole-v1 environment with Proximal Policy Optimization (PPO), using \texttt{SimplyQRL}~\cite{lazaro_simplyqrl_2025}, a previously validated framework specifically designed to ensure consistent comparisons between quantum and classical agents without altering the underlying RL pipeline. Simulations are carried out on PennyLane’s statevector backend in a noiseless setting, focusing the analysis on algorithmic and architectural effects rather than hardware noise. By adopting this controlled setup, we are able to evaluate the influence of each block under fair, reproducible, and directly comparable conditions.

Our results are organised as a cascade that reflects the dependency structure between components. First, we analyse post-PQC inference through the conceptually simple yet underexplored Output Reuse (OR) technique, showing that although it is a classical mechanism, it exhibits differential efficacy in classical versus hybrid pipelines, with gains that depend on the surrounding circuit structure. Second, we investigate observation embeddings with a focus on Data Reuploading (DR) and show that advanced embeddings are associated with improved trainability and stability, displaying patterns consistent with the mitigation of Barren Plateaus (BPs). Notably, even within the same angle-embedding family, distinct embedding philosophies induce predictably scalable yet qualitatively different learning behaviours under increasing DR depth, underscoring the sensitivity of hybrid performance to implementation details. Third, we study ansatz design and entanglement, observing that small architectural changes can materially affect trainability and stability, in some cases offsetting the benefits of classical post-processing. Together, these results highlight the nuanced interplay between classical and quantum contributions and provide controlled empirical evidence to guide the future design of QRL architectures.

\paragraph{Main findings and contributions.}
This work presents, to our knowledge, the first systematic decomposition of PQC-based hybrid QRL pipelines that isolates and evaluates the contribution of their three core components: post-PQC inference, observation embedding, and ansatz design, within a unified and reproducible framework. This controlled analysis clarifies how hybrid performance arises from the structured interaction between quantum and classical subsystems rather than from isolated design choices.

(i) \emph{Post-PQC inference:} We expand on the original study of Output Reuse (OR)~\cite{hsiao_unentangled_2022}, providing new empirical evidence that its benefits are specific to PQC-based agents, while purely classical counterparts exhibit marginal or unstable improvements. This confirms that OR’s effectiveness depends on meaningful quantum representations, highlighting its role as a genuine hybrid interaction mechanism rather than a classical scaling heuristic.

(ii) \emph{Observation embeddings:} We show that even within the same angle-encoding family, differing application philosophies substantially modify the impact of advanced techniques such as Data Reuploading (DR), either limiting or enhancing circuit trainability as depth (layers) and width (qubits) increase.

(iii) \emph{Ansatz design:} A targeted entanglement ablation reveals that complex hybrid pipelines are highly sensitive to architectural changes, an aspect questioned in prior studies~\cite{kolle_study_2024}. The variability observed across entanglement configurations suggests that task-specific ansatz families may represent a promising avenue for future QRL research, complementing mechanisms such as DR and post-PQC processing.

In summary, we contribute (a) a controlled, component-wise protocol that disentangles quantum and classical effects under fixed PPO–CartPole conditions; (b) extended classical controls enabling causal attribution of hybrid gains to genuine quantum–classical synergy; and (c) a reproducible benchmarking framework and open dataset supporting direct re-evaluation and extension. These findings provide essential empirical evidence of the interplay between quantum and classical components in the rapidly expanding landscape of hybrid QRL architectures, shaped by the practical constraints of current NISQ devices.

\section{Related Work}
\label{sec:related_work}

Research on QRL spans from early theoretical proposals to the current predominance of PQC-based methods. Initial formulations of QRL were largely theoretical, motivated by potential speed-ups via quantum search and unitary evolution \cite{dong_quantum_2008}. These works, together with subsequent quantum-inspired approaches, established conceptual links between quantum information processing and classical RL but offered limited empirical validation. The field shifted decisively with the advent of PQCs as machine learning models \cite{benedetti_parameterized_2019,schuld_quantum_2019,cerezo_variational_2021}, which provided a practical route to hybrid agents trainable by gradient descent. PQCs also share strong structural analogies with classical NNs \cite{perez-salinas_data_2020}, further increasing interest in testing them on domains where NNs have already demonstrated success, such as RL benchmarks. Beyond PQCs, alternative directions include Hamiltonian-based formulations for combinatorial optimisation \cite{kruse_hamiltonian-based_2024}, compression and scheduling oriented training frameworks \cite{liu_quantum-train_2024,liu_qtrl_2024}, continuous-action formulations \cite{wu_quantum_2025}, and advances in policy/action decoding \cite{meyer_quantum_2023}; while valuable, these lines currently have less standardised empirical coverage than PQC-centric hybrids in canonical control environments.

Recent surveys synthesise this transition and emphasise persistent challenges around trainability, benchmarking, and robustness \cite{meyer_survey_2024,alomari_survey_2025}. Within the PQC-based paradigm, hybrid agents have been instantiated across standard RL families: replacing the Q-network in DQN and scaling to Atari-style tasks \cite{lockwood_playing_2021,skolik_quantum_2022,freinberger_quantum-classical_2024}, adapting policy-gradient methods \cite{sequeira_policy_2023}, and developing actor–critic variants such as A2C, PPO, and SAC \cite{kolle_quantum_2024,kolle_study_2024,lan_variational_2021,acuto_variational_2022}. Despite their architectural diversity, these models consistently follow a three-block pipeline: (i) an embedding of observations into quantum states, (ii) a PQC ansatz (with or without entanglement), and (iii) a post-measurement inference layer mapping expectation values to policies or value estimates. However, most studies vary multiple components simultaneously, which complicates attribution of gains to specific blocks.

\subsection{Observation embeddings and Data Reuploading}

The embedding block is a fundamental component of PQCs and, by extension, of PQC-based QRL. In quantum machine learning (QML), data encoding is recognised as decisive for both expressibility and trainability, and remains an open research challenge \cite{schuld_quantum_2019,cerezo_variational_2021,schuld_effect_2021,khan_beyond_2024}. In QRL, the challenge is compounded by the size and diversity of observation spaces, which often require normalisation or preprocessing before encoding. Most approaches therefore combine a normalisation step with an encoding scheme that ensures distinct observations are mapped to distinct quantum states, thereby avoiding representational ambiguities \cite{kolle_quantum_2024,skolik_quantum_2022,hsiao_unentangled_2022}. While such strategies have enabled practical QRL models, they are often difficult to train in practice, as reported in early implementations \cite{chen_variational_2020,skolik_quantum_2022}.

Within this context, \emph{Data Reuploading} (DR) was introduced in QML as a \emph{Universal Quantum Classifier} (UQC), where repeated encodings mimicked the hidden layers of classical neural networks \cite{perez-salinas_data_2020}. It was later adopted in QRL, with Skolik et al. providing the first effective demonstration \cite{skolik_quantum_2022}, followed by applications in DQN and PPO-based agents \cite{coelho_vqc-based_2024,kolle_study_2024,hohenfeld_quantum_2024}. DR is generally regarded as a way to enrich representations and stabilise training, with some evidence that it may mitigate BPs \cite{holmes_connecting_2022}. At the same time, its evaluation has rarely been isolated: DR is typically studied alongside changes in ansatz structure, preprocessing, or inference layers, and even stronger works such as Coelho et al. restrict analysis to a single ansatz type \cite{coelho_vqc-based_2024}. Moreover, recent work suggests that the impact of stacking multiple DR layers is still analytically unclear \cite{alamo_is_2025}. As a result, while DR has become almost a default choice in QRL pipelines, systematic studies across different ansatz configurations are still missing.

% (Optional) mini esquema de DR

\subsection{Ansatz design and entanglement}

Ansatz design is a long-standing challenge in QML, as it directly impacts expressibility, trainability, and the onset of BPs \cite{schuld_circuit-centric_2020,cerezo_variational_2021}. Prior studies have shown that the choice of gates, connectivity, and depth governs whether a model can approximate the target function and maintain optimisation feasible \cite{cerezo_cost_2021,nakaji_expressibility_2021,schuld_effect_2021}. In particular, highly expressive ansätze, such as those approaching unitary 2-designs, may approximate rich function classes but also suffer from exponentially vanishing gradients \cite{holmes_connecting_2022,sequeira_trainability_2024}. Within this line, layers spanning the single-qubit $SU(2)$ space have become standard building blocks for QML circuits. A generic layer of this kind on $n$ qubits can be expressed compactly as

\begin{equation}
U_{\text{layer}}(\boldsymbol{\Theta})
=
\mathcal{E}\;
\Bigg(\;\bigotimes_{j=1}^{n}
R_{Z}\!\big(\theta^{(3)}_{j}\big)\,
R_{Y}\!\big(\theta^{(2)}_{j}\big)\,
R_{Z}\!\big(\theta^{(1)}_{j}\big)
\Bigg)
\label{eq:su2-layer}
\end{equation}

where the tensor product supplies per-qubit $SU(2)$ rotations and the
entangling layer
\[
\mathcal{E}=\prod_{(a,b)\in E} G_{ab}
\]
acts over edges $E$ of a chosen connectivity (e.g., line or ring) with
two-qubit gate $G_{ab}$ (such as $CZ$, $CX$, or controlled rotations).
Repetition over $L$ layers yields

\begin{equation}
U(\boldsymbol{\Theta})=\prod_{\ell=1}^{L} U_{\text{layer}}^{(\ell)}(\boldsymbol{\Theta}^{(\ell)})
\end{equation}

This formulation captures the general pattern adopted in most hybrid models. Practical variants arise by altering the rotation order or the gate set, changing $\mathcal{E}$, or adapting the connectivity to the underlying hardware topology. In QRL, this general family of ansätze underpins many circuit architectures, differing mainly in depth, rotation order, and entangling scheme. Several studies employ modifications of this template tailored to specific algorithms or hardware constraints \cite{skolik_quantum_2022,jerbi_parametrized_2021,chen_variational_2020,lockwood_playing_2021,freinberger_quantum-classical_2024,coelho_vqc-based_2024,hohenfeld_quantum_2024}. Despite implementation differences, these models share the same structural elements: an angle or basis embedding, parameterised single-qubit rotations, an entangling layer, and a final measurement of all qubits.

As in QML, no consensus exists in QRL regarding the most suitable ansatz or entanglement scheme. Entanglement can enhance expressibility but also hinder trainability, with some studies showing that even unentangled circuits may perform competitively \cite{hsiao_unentangled_2022}. Systematic analyses, such as Kölle et al. \cite{kolle_study_2024}, suggest that ansatz design may exert less influence than techniques like DR or learning-rate scheduling. This ambiguity leaves open whether ansatz structure significantly affects BPs, expressibility, or optimisation in hybrid QRL pipelines.

\subsection{Post-PQC inference and Output Reuse}

The post-processing block is the final stage of PQC-based QRL pipelines, responsible for transforming quantum measurement outputs into representations compatible with classical RL algorithms such as DQN or PPO. Early works relied on simple strategies, including single-qubit measurement combined with a classical interpretation function \cite{chen_variational_2020}, or output scaling (OS) mechanisms to map expectation values into suitable ranges for Q-value approximation in DQN \cite{skolik_quantum_2022,freinberger_quantum-classical_2024}. More advanced approaches have also been explored, including action/policy decoding schemes \cite{meyer_quantum_2023} and continuous-action formulations \cite{wu_quantum_2025}. 

A common technique across this literature is the use of classical NNs as interpretation layers, mapping quantum expectation values to the logits or value functions expected by RL algorithms. This practice is widespread across policy-gradient and actor–critic methods \cite{lan_variational_2021,acuto_variational_2022,kolle_quantum_2024,freinberger_quantum-classical_2024}, and serves a dual role: it implicitly rescales quantum outputs (an implicit form of OS) while enabling flexibility with respect to the number of qubits, observation sizes, and action spaces. This design allows PQCs to be combined with diverse environments, from \textit{FrozenLake} and \textit{CartPole} to \textit{Acrobot}, \textit{LunarLander}, and robotics navigation tasks.

\begin{wrapfigure}{r}{0.475\textwidth}
  \centering
  \vspace{-1em}
  \includegraphics[width=0.4\textwidth]{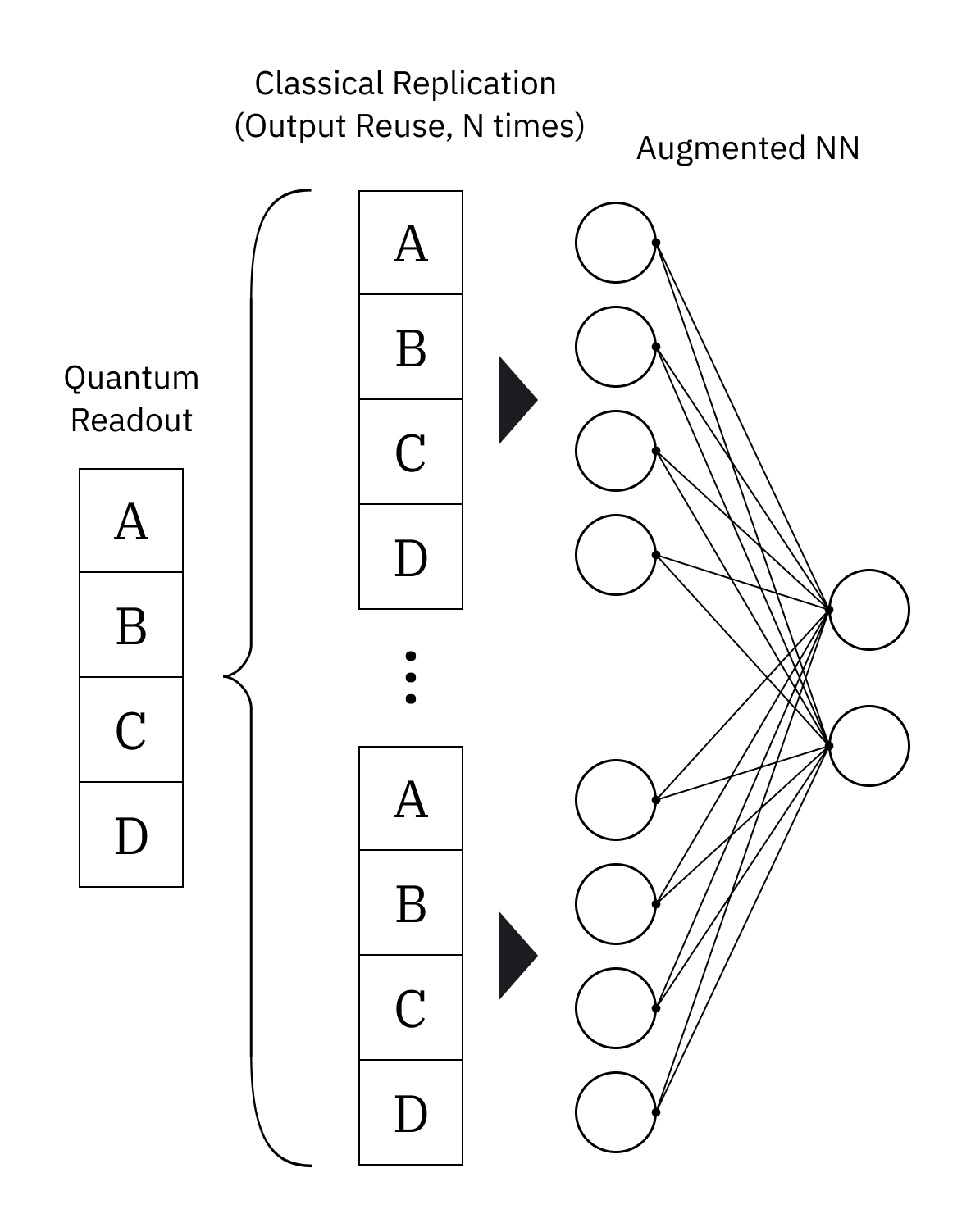}
  \vspace{-0.5em}
  \caption{Illustration of the Output Reuse (OR) technique~\cite{hsiao_unentangled_2022}. The PQC output vector (A–D) is replicated $N$ times before the classical interpretation layer, increasing the input dimensionality of the neural head.}
  \vspace{-1em}
  \label{fig:or-concept}
\end{wrapfigure}

Within this category, Hsiao et al. introduced \emph{Output Reuse} (OR), a technique that replicates the quantum output tensor $N$ times before the classical interpretation layer, thereby enlarging the effective input dimension of the NN  (see Fig.~\ref{fig:or-concept}). Although OR is purely classical, their results showed that it could enhance the learning performance of PQC-based agents, particularly when combined with specific circuit structures. This raises an important attribution issue: once additional NN-based processing or tricks such as OR and learnable OS are introduced, it becomes unclear which part of the pipeline is actually responsible for learning. Existing works, including those of Hsiao et al. \cite{hsiao_unentangled_2022} and Kölle et al. \cite{kolle_quantum_2024}, acknowledge this difficulty but do not provide systematic analyses of the post-processing block in isolation.

\subsection{Cross-cutting limitations and benchmarking needs}

Across embeddings, ansatz design, and post-PQC inference, prior work often varies multiple components simultaneously, making it difficult to attribute performance gains to any single block \cite{hsiao_unentangled_2022,skolik_quantum_2022,coelho_vqc-based_2024,sequeira_trainability_2024,kolle_study_2024}. This conflation obscures whether improvements originate from intrinsic circuit properties, classical post-processing, or algorithmic heuristics. Sensitivity to circuit depth, entanglement, and optimisation settings further complicates interpretation, with instability and vanishing-gradient phenomena frequently reported in QRL studies~\cite{sequeira_trainability_2024,kolle_study_2024}. These weaknesses are partly structural, as QRL is a young field still lacking shared benchmarks and evaluation standards, and partly methodological, as the absence of systematic ablations prevents disentangling quantum from classical contributions.

Recent efforts have begun to address this gap by proposing statistically robust benchmarking protocols for QRL. In particular, Meyer et al. \cite{meyer_benchmarking_2025} advocate for reproducible evaluation based on statistically significant differences in sample complexity, while Kruse et al. \cite{kruse_benchmarking_2025} call for standardised protocols to mitigate the variance introduced by environment choice, seeding, and reporting conventions. Building on this momentum, our work introduces a controlled, component-wise analysis of PQC-based QRL pipelines. Using a unified PPO protocol in the canonical \textit{CartPole} environment, we systematically isolate each major block: embeddings with DR, ansatz design with and without entanglement, and post-PQC inference with OR. All experiments are implemented within \texttt{SimplyQRL}, a framework specifically designed to dissect hybrid QRL models while ensuring algorithmic consistency across different configurations \cite{lazaro_simplyqrl_2025}. By holding algorithm, environment, and evaluation budget fixed, and by adopting learning curves (average episodic reward vs. training steps) as our central metric, we capture sample efficiency, trainability, and the onset of BPs, which motivates a closer examination of these circuit properties in the following methodological section.

\section{Methodology}
\label{sec:methodology}

We now describe the methodological framework underlying our study, designed to isolate the effect of each pipeline block, under a unified PPO–CartPole setting that ensures consistency and reproducibility across all experiments. Before detailing our experimental design, it is important to clarify two key properties that will be referenced throughout this work: \textit{trainability} and \textit{expressibility}. 

In the context of PQCs, trainability refers to the ability of a circuit to be efficiently optimised, which is directly influenced by the presence or absence of BPs, regions of the parameter space where gradients vanish and learning stagnates~\cite{holmes_connecting_2022,sequeira_trainability_2024,alamo_is_2025}. Expressibility, in turn, denotes the extent to which a PQC can explore the underlying Hilbert space, determining the diversity and representational capacity of the quantum states it generates~\cite{nakaji_expressibility_2021,holmes_connecting_2022,alamo_is_2025}.

In this study, we assess trainability empirically through learning dynamics (average episodic reward vs.\ training steps), while expressibility is discussed qualitatively through architectural proxies: circuit depth, qubit count, and entanglement pattern, which have been shown to correlate with expressibility in prior work. These notions frame our methodological design and provide the conceptual basis for interpreting the results that follow.

\subsection{Design principles and global set-up}

Our methodological design follows the core ideas introduced in Sections~\ref{sec:intro}--\ref{sec:related_work}: PQC-based QRL pipelines comprise three tightly coupled blocks, post-PQC inference, observation embedding, and ansatz. To disentangle their individual effects, we adopt a controlled \emph{block-wise} evaluation in which the RL algorithm, environment, training budget, and evaluation protocol remain fixed across all experiments.

Unless otherwise stated, all agents are trained with PPO on CartPole-v1 for 100\,k environment steps per run, using ten fixed seeds. This configuration supports comparability with prior QRL studies while keeping variance manageable and compute affordable~\cite{jerbi_parametrized_2021,skolik_quantum_2022,sequeira_policy_2023,kolle_study_2024,coelho_vqc-based_2024}. Learning curves (average episodic return vs.\ environment steps) are reported as the mean with a shaded standard-deviation band across seeds, reflecting sample efficiency, stability, and final performance.

\subsection{Benchmarking framework: \texttt{SimplyQRL}}
\label{subsec:simplyqrl}

All experiments are implemented in \texttt{SimplyQRL}, a lightweight benchmarking framework that preserves the \emph{core} RL algorithm and training loop while making the agent architecture (classical, quantum, or hybrid) fully configurable via declarative dictionaries. In this set-up, the only moving parts are the three pipeline blocks: (i) the observation embedding, (ii) the PQC ansatz (including entanglement), and (iii) the post-measurement inference layer. PPO and DQN baselines were adapted from \texttt{CleanRL} and verified for functional equivalence under identical hyperparameters; PQCs are implemented as \texttt{TorchLayer}s via PennyLane, with explicit attention to batched execution for efficiency. A detailed description of the framework, validation steps, and configuration interface appears in the accompanying dissertation \cite{lazaro_simplyqrl_2025}.

\subsection{Common training protocol and controls}
\label{subsec:protocol}

To isolate the contribution of each block, we enforce the following uniform protocol:

\begin{itemize}
\item \textbf{Environment and budget.} All runs use CartPole-v1 for 100\,k steps with 10 seeds. The PPO implementation, optimiser, and rollout/evaluation schedules are held fixed across experiments.\footnote{Full PPO hyperparameters and training scripts are available in the project repository.}
\item \textbf{Observation preprocessing.} For each ansatz template we apply the observation normalisation and preprocessing prescribed in the corresponding prior work to avoid adding confounding factors.
\item \textbf{Measurement and inference.} PQCs are measured as Z-axis expectation values on all qubits. These expectations feed a single linear layer that maps directly to action logits (actor) or a scalar value (critic). No hidden layers are used in the default inference block; when OR is active, only the input dimensionality grows, keeping the mapping linear.
\item \textbf{Actor--critic symmetry.} Actor and critic use separate but architecturally identical PQCs and inference heads; the critic outputs a single value while the actor outputs two logits (one per action).
\item \textbf{Quantum backend.} All quantum simulations use PennyLane’s  state-vector backend (\texttt{lightning.qubit}) in a noiseless, no-shot setting. Gradients are computed with adjoint differentiation, which provides exact expectation-value derivatives without stochastic sampling.
\end{itemize}

\subsection{Experimental factors and designs}
\label{subsec:designs}

We organise experiments around the three blocks, with minimal changes elsewhere. Each block is examined through a tailored experimental strategy suited to its role in the hybrid pipeline. For the post-PQC inference stage, we vary the number of output replications (R) and include a fully classical control to disentangle the quantum contribution. For the embedding block, we contrast two distinct angle-embedding philosophies and assess their scaling behaviour by increasing the number of DR layers. Finally, for the ansatz, we conduct a targeted ablation of entanglement, toggling its presence within otherwise identical architectures. These complementary strategies ensure that each component is evaluated under conditions that are both comparable and representative of its functional behaviour within the unified PPO–CartPole framework.

\paragraph{(A) Post-PQC inference: Output Reuse (OR).}
Following Hsiao et~al., we realise OR by concatenating $R$ replicas of the PQC output vector immediately before the linear head ($R \in \{4,8,16,32\}$) \cite{hsiao_unentangled_2022}. As OR is purely classical, we test whether its benefits depend on the presence of a quantum block or obtain equally in a fully classical agent. Accordingly, we include a classical control that applies the same OR operation to the observation vector (optionally reduced for dimensional parity) and uses an identical linear readout. 

This design equalises input dimensionality and yields a similar parameter count across hybrid and classical variants (the hybrid retains a small, fixed set of PQC parameters that does not scale with $N$), thereby isolating the PQC contribution.

\begin{figure}[t]
  \centering
  \includegraphics[width=\textwidth]{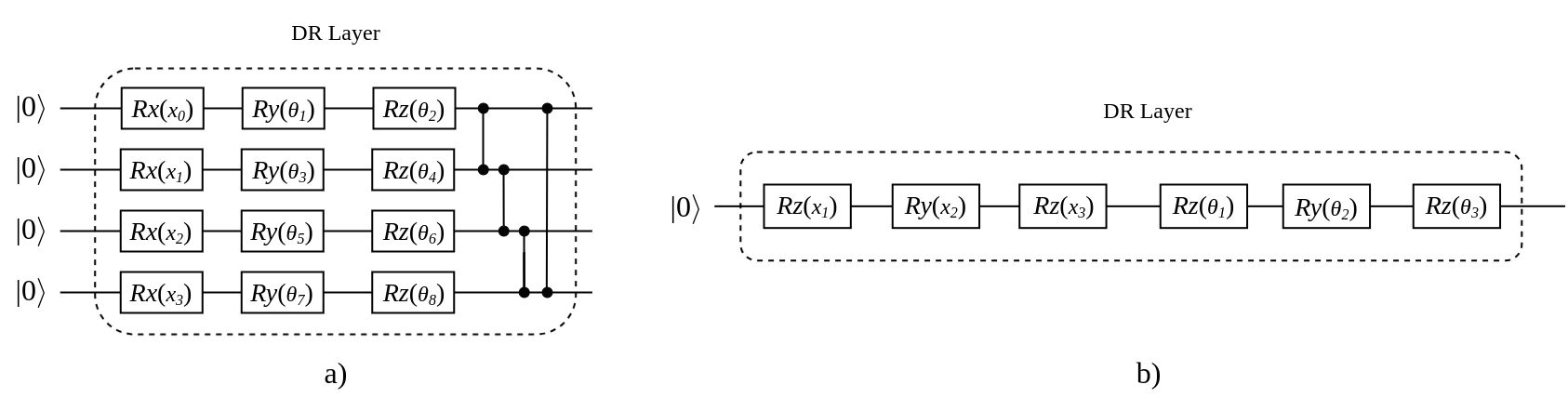}
  \caption{Comparison of the two angle-embedding philosophies evaluated. (a) Skolik-style embedding encodes one feature per qubit through a single $R_X(x)$ rotation.(b) UQC-style embedding follows a $R_ZR_YR_Z$ pattern that encodes multiple data components per qubit and naturally extends to DR~\cite{perez-salinas_data_2020}. The dotted area marks one DR layer, whose repetition defines circuit depth L. This distinction underlies our analysis of embedding expressibility and scalability.}
  \label{fig:emb-configs}
\end{figure}

\paragraph{(B) Observation embedding and Data Reuploading (DR).}
The second set of experiments evaluates how distinct embeddings affect RL performance and how they scale under DR. We contrast two angle-embedding templates (Fig.~\ref{fig:emb-configs}) that inject information very differently while both being $SU(2)$-complete per qubit:

\begin{enumerate}
\item \textbf{Skolik-style \cite{skolik_quantum_2022}}: a simple per-qubit embedding $R_X(x)$ (one datum per qubit) paired with a shallow hardware-efficient ansatz; in practice, $N$ qubits are used for $N$ observation features.
\item \textbf{UQC-style \cite{perez-salinas_data_2020}}: a layered per-qubit embedding $R_Z(x_1)R_Y(x_2)R_Z(x_3)$, allowing three data per qubit at the cost of increased depth; the parametrised per-qubit layer mirrors the same Z–Y–Z factorisation.
\end{enumerate}

% --- FIGURA DE LOS DOS TIPOS DE EMBEDINGS USADOS ---

Our comparative factor is the \emph{embedding philosophy}. Although embedding and ansatz cannot be fully decoupled, the $SU(2)$-complete constructions support a meaningful parallelism. We therefore evaluate each template in its base configuration and under depth scaling via DR, and we assess sensitivity to the number of available qubits by contrasting one-feature-per-qubit against multi-feature-per-qubit regimes. This yields a controlled comparison of embedding strategies for CartPole, clarifying their impact on trainability and informing PQC-based QRL design.

\begin{figure}[t]
  \centering
  \includegraphics[width=\textwidth]{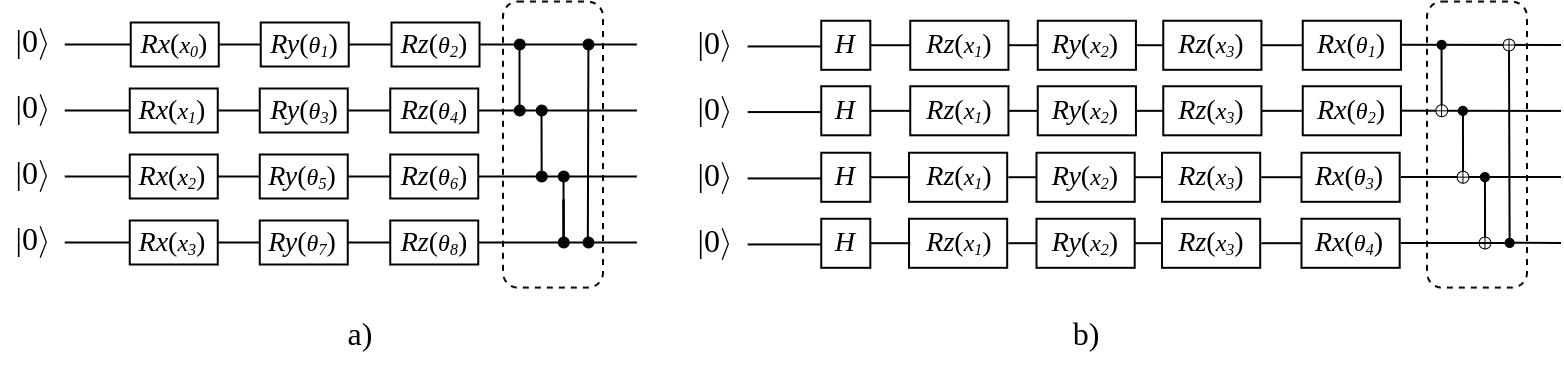}
  \caption{Circuit templates used in the entanglement ablation. (a) Template A (Skolik-derived) uses a ring of CZ gates to introduce entanglement between adjacent qubits. (b) Template B (Hsiao-derived) is initially unentangled and later extended with a ring of CNOTs. The dotted region denotes the entanglement layer toggled on/off in the experiments. This design isolates the influence of entanglement under otherwise identical conditions.}
  \label{fig:ent-configs}
\end{figure}

\paragraph{(C) Ansatz design and entanglement.}
A comprehensive ansatz search is beyond scope and would require a highly granular study over rotations, connectivity, depth, and qubit count (see automated explorations in \cite{wang_automated_2023}). Instead, we perform a \emph{targeted entanglement ablation} on two literature-grounded templates with opposing default assumptions about connectivity (see Fig.~\ref{fig:ent-configs}). For each template we evaluate a \emph{within-template} pair that differs only by the presence of an entangling layer. 

\begin{itemize}
\item \textbf{Template A (Skolik-derived):} typically instantiated with a ring of CZ gates. We compare the standard configuration against an identical variant with entanglement removed.
\item \textbf{Template B (Hsiao-derived):} originally implemented without entanglement. We compare the standard configuration against a matched variant with a ring of CNOTs inserted between per-qubit rotation blocks.
\end{itemize}

This design isolates \emph{entanglement as a single structural toggle} under otherwise identical conditions. Within each template and condition, we further vary DR depth (Templates A \& B) and OR repetitions (Template B) to examine whether their effectiveness changes due to the presence or absence of entanglement. In doing so, we obtain a targeted quantification of the effect of entanglement on trainability and stability in two concrete, literature-grounded cases, and we directly test whether any gains (or degradations) are template-agnostic. This focused ablation also speaks to recent claims that ansatz design may be secondary to techniques such as DR \cite{kolle_study_2024}, by isolating entanglement as a specific architectural factor within the ansatz.

% --- FIGURA DE LOS DOS ANSATZ ENTRELAZADOS ---

\subsection{Evaluation and presentation of results}
\label{subsec:metrics}

We report learning curves (episodic return vs.\ environment steps) as the mean curve with a shaded $\pm$1 standard deviation band across ten seeds. This is the standard summary in RL and predominates in recent QRL studies, enabling direct comparison with prior work \cite{hsiao_unentangled_2022,skolik_quantum_2022,coelho_vqc-based_2024,freinberger_quantum-classical_2024,kolle_study_2024}. In our setting, these learning dynamics directly reflect the trainability of each agent, with early stagnation or high across-seed variance indicating reduced optimisation performance and potential barren-plateau effects. While expressibility is not evaluated analytically, we interpret architectural trends qualitatively through known structural factors: circuit depth, number of qubits, and entanglement pattern, which correlate with expressibility in PQCs \cite{nakaji_expressibility_2021,holmes_connecting_2022}. 

While alternative, statistically stricter benchmarking protocols have been proposed \cite{meyer_benchmarking_2025,kruse_benchmarking_2025}, we prioritise consistency with the evaluation practice of the works we directly compare against \cite{hsiao_unentangled_2022,kolle_study_2024,coelho_vqc-based_2024}.

Results are presented in an order that reflects the dependency structure of the hybrid pipeline and facilitates cumulative interpretation. We begin with \textit{post-PQC inference (OR)}, whose behaviour can mask or amplify downstream effects and for which classical controls are available. We then examine \textit{embedding and DR}, where trainability and signs of BP mitigation depend on the interaction between embeddings and ansatz design. Finally, we analyse \textit{ansatz and entanglement}, using the preceding evidence to contextualise when additional expressibility enhances or hinders optimisation in hybrid agents.

\section{Results}
\label{sec:results}

We now present the results obtained from the experiments described in Section \ref{sec:methodology}. The order of presentation follows the dependency structure outlined in the methodology: first post-PQC inference with Output Reuse, then observation embedding and Data Reuploading, and finally ansatz design and entanglement. This progression allows each block to be interpreted in the context of the preceding analyses.

\subsection{Post-PQC Inference: Output Reuse}
\label{subsec:inference}

\begin{figure*}[t]
  \centering
  % --- Subfigura (a) ---
  \begin{subfigure}[t]{0.495\textwidth}
    \centering
    \includegraphics[width=\linewidth]{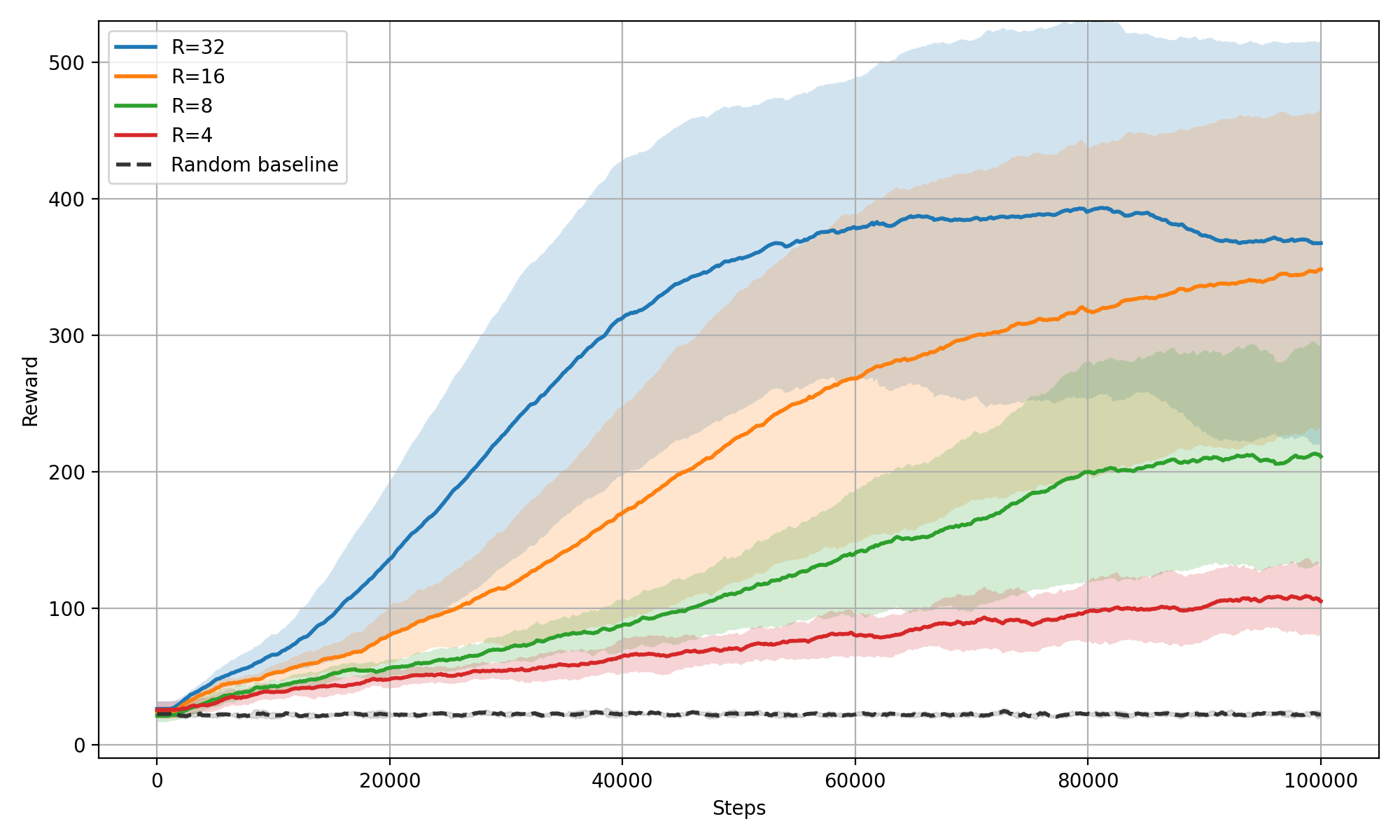}
    \caption{Hybrid agent followed by OR.}
    \label{fig:or-results-quantum}
  \end{subfigure}
  \hfill
  % --- Subfigura (b) ---
  \begin{subfigure}[t]{0.495\textwidth}
    \centering
    \includegraphics[width=\linewidth]{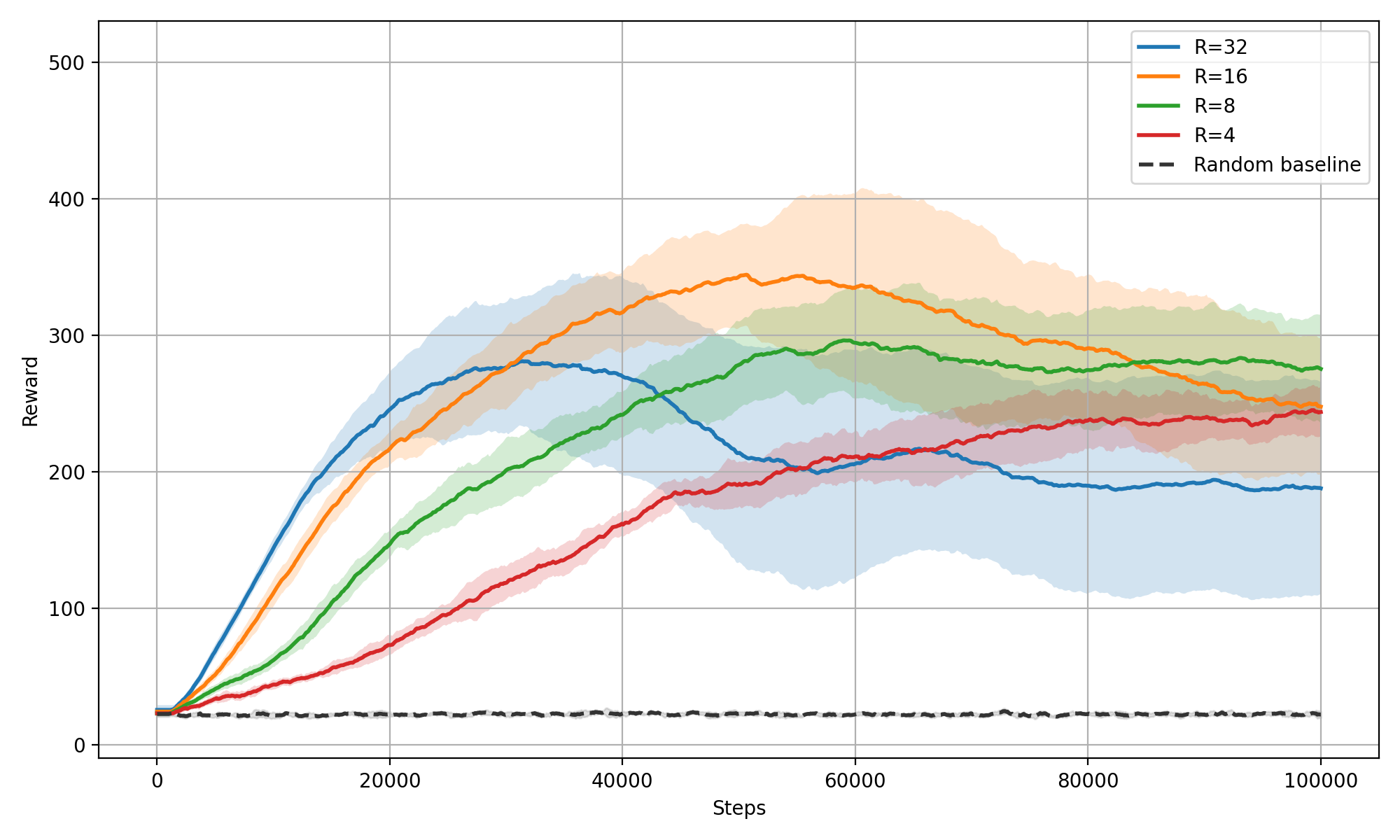}
    \caption{Classical agent with observation preprocessing and OR.}
    \label{fig:or-results-classical}
  \end{subfigure}
  % --- Pie general ---
  \caption{Learning curves for hybrid and classical agents under increasing Output Reuse ($R$). While OR consistently improves hybrid performance, classical agents show only minor or unstable gains, confirming that OR’s efficacy depends on the presence of a quantum block.}
  \label{fig:or-results}
\end{figure*}

Fig.~\ref{fig:or-results} compares the behaviour of hybrid (quantum) and purely classical agents when subjected to increasing levels of OR, corresponding to the number of replicated quantum readouts $R$. Fig.~\ref{fig:or-results-quantum} shows results for the hybrid configuration following the Hsiao \emph{et~al.}~\cite{hsiao_unentangled_2022} architecture, while Fig.~\ref{fig:or-results-classical} reports the equivalent experiment for a classical control agent where the observation vector is pre-processed to mimic the hybrid embedding step.

For hybrid agents employing OR, a clear improvement emerges as the reuse factor $R$ increases  at the PQC output. Performance rises consistently between $R=4$, $8$, and $16$, with $R=32$ beginning to exhibit diminishing returns accompanied by greater seed-to-seed variability, likely due to over-parameterisation of the inference head relative to the PQC block.

When the same OR procedure is applied to a purely classical agent (without a PQC), the pattern diverges notably. Improvements with increasing $R$ are far less consistent, with mild gains up to $R=4$ and $8$, followed by a marked degradation in learning stability for larger values of $R$. 

These results suggest that the quantum measurement vector at the end of the PQC encodes meaningful information aligned with the agent’s policy, yet suffers from under-dimensionality and potential scale mismatch between quantum expectations and the downstream logits or value estimates, as seen in \cite{skolik_quantum_2022,freinberger_quantum-classical_2024}. In this context, OR appears to alleviate such limitations by effectively amplifying the representational capacity of the inference layer. However, it remains unclear whether this effect generalises to other PQC architectures or is specific to the design introduced by Hsiao \emph{et~al.}~\cite{hsiao_unentangled_2022}.

\subsection{Observation Embedding and Data Reuploading}
\label{subsec:embedding}

Transitioning from inference to observation embeddings, we compare two distinct philosophies of angle embedding (Fig.~\ref{fig:emb-configs}) under varying numbers of DR layers $L$ and qubits $Q$.

\begin{figure*}[t]
  \centering
  % --- Subfigura (a) ---
  \begin{subfigure}[t]{0.495\textwidth}
    \centering
    \includegraphics[width=\linewidth]{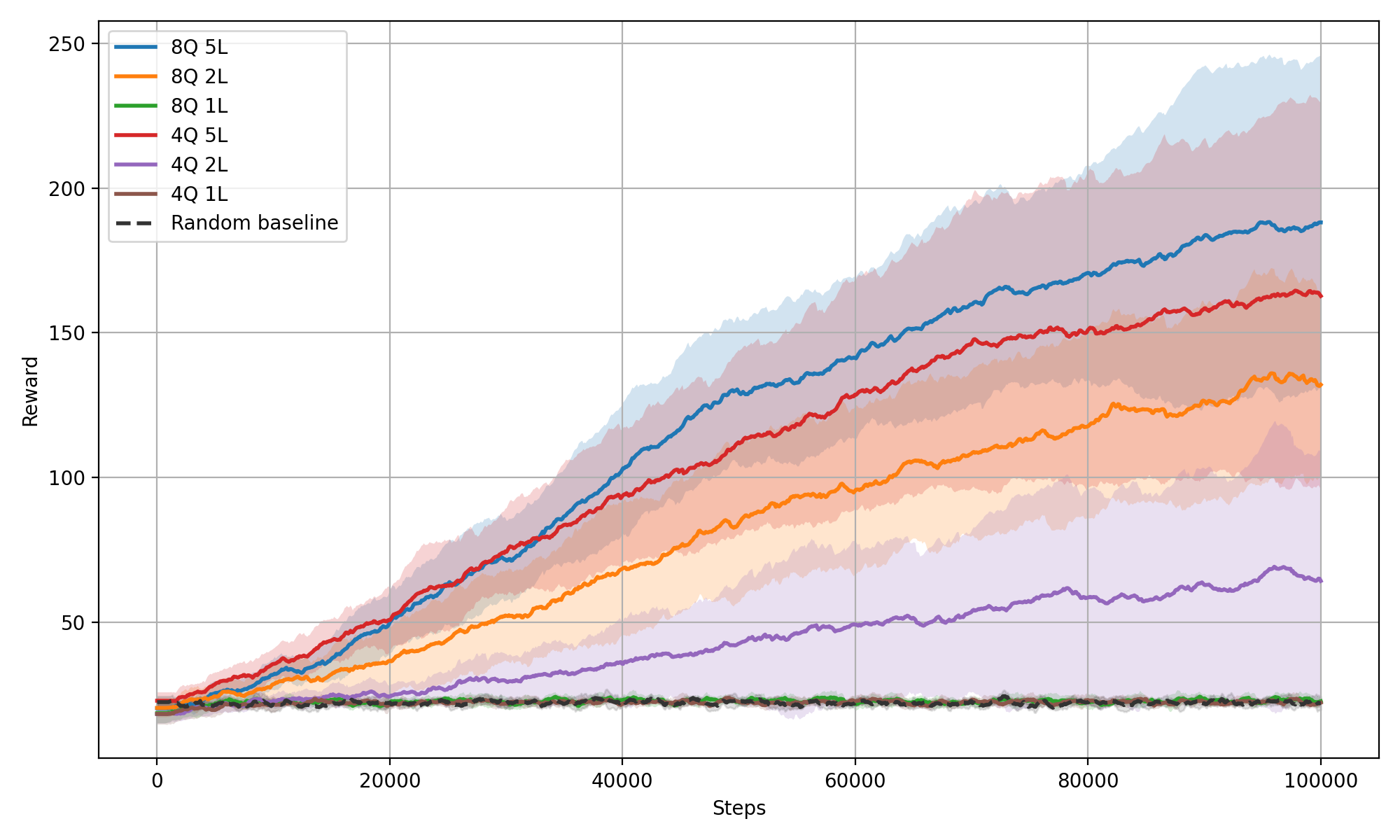}
    \caption{Skolik-style embedding.}
    \label{fig:dr-results-skolik}
  \end{subfigure}
  \hfill
  % --- Subfigura (b) ---
  \begin{subfigure}[t]{0.495\textwidth}
    \centering
    \includegraphics[width=\linewidth]{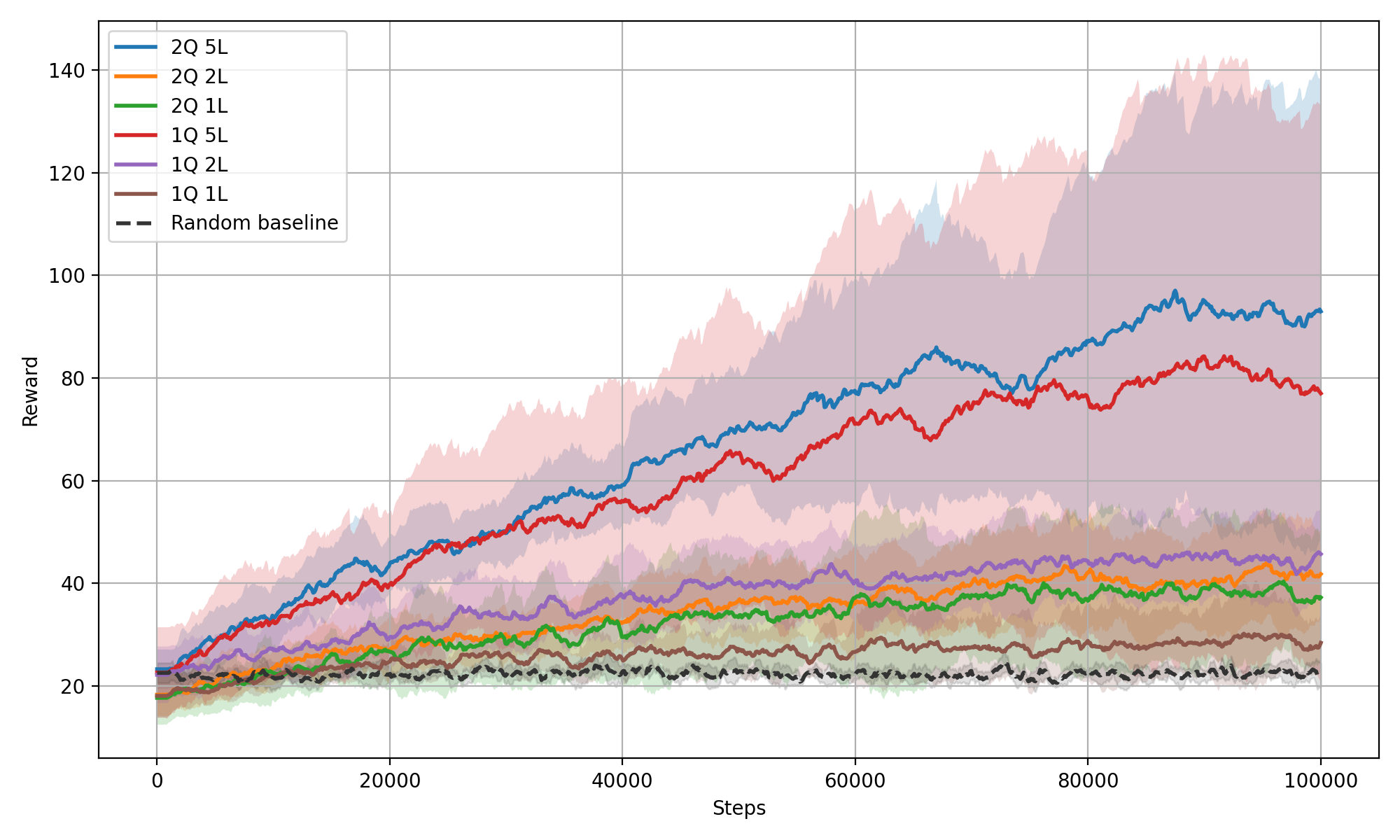}
    \caption{UQC-style embedding.}
    \label{fig:dr-results-uqc}
  \end{subfigure}
  % --- Pie general ---
  \caption{Learning curves for the two embedding strategies under varying DR depth ($L$) and qubit count ($Q$). Both exhibit predictable scaling with $L$ and $Q$, though the UQC-style embedding shows higher base trainability at low depth.}
  \label{fig:dr-results}
\end{figure*}

Fig.~\ref{fig:dr-results-skolik} shows results for the Skolik-style embedding. Agents with a single DR layer ($L=1$) fail to learn effectively, regardless of the number of qubits. Since all outputs are rescaled through an identical linear inference layer, this behaviour cannot be attributed to output scaling; rather, the PQC itself exhibits poor trainability within the given budget, suggesting that its parameterisation or depth is insufficient for stable optimisation. When increasing depth beyond $L=2$, the circuits become expressible enough to achieve stable learning, showing marked performance improvements up to $L=5$. This trend aligns with the notion that deeper parameterised layers enhance circuit expressibility, thereby improving trainability by enabling more efficient optimisation \cite{nakaji_expressibility_2021,holmes_connecting_2022}. The same pattern extends across qubit counts: the expanded eight-qubit configuration achieves stronger gains as depth increases, suggesting that this embedding scales positively both in depth and width.

Fig.~\ref{fig:dr-results-uqc} presents the equivalent experiment for the UQC-style embedding. In this configuration, information is compacted into a single qubit, which requires deeper reuploading to reach comparable results. Noticeably, even the shallowest configuration ($L=1$) exhibits some reward improvement over time, indicating that the base embedding is inherently more trainable, yet potentially less expressible, than its prior counterpart. Increasing either depth or qubit count further enhances performance, showing that both embedding families maintain scalability in these two dimensions.

Overall, both embeddings display broadly parallel trends: performance improves systematically as DR depth or qubit width increases, consistent with the expected benefits of reuploading architectures. This predictable scaling behaviour partly explains the widespread adoption of DR in QRL research. Nonetheless, meaningful differences remain, particularly in base trainability and representational capacity, even though both methods employ angle embeddings. Prior work has highlighted these distinctions in broader QML contexts \cite{casas_multidimensional_2023}, yet within QRL they remain underexplored. Our findings therefore indicate that not only the choice of embedding type but also its structural design is a critical factor in ensuring robust performance in hybrid QRL models.

\subsection{Ansatz design: Entanglement}
\label{subsec:entanglement}

We now analyse the role of entanglement as a design element within the ansatz. Results are presented for two ansatz templates: Template~A (Skolik-derived) and Template~B (Hsiao-derived). 

\begin{figure}[t]
  \centering
  \includegraphics[width=0.95\linewidth]{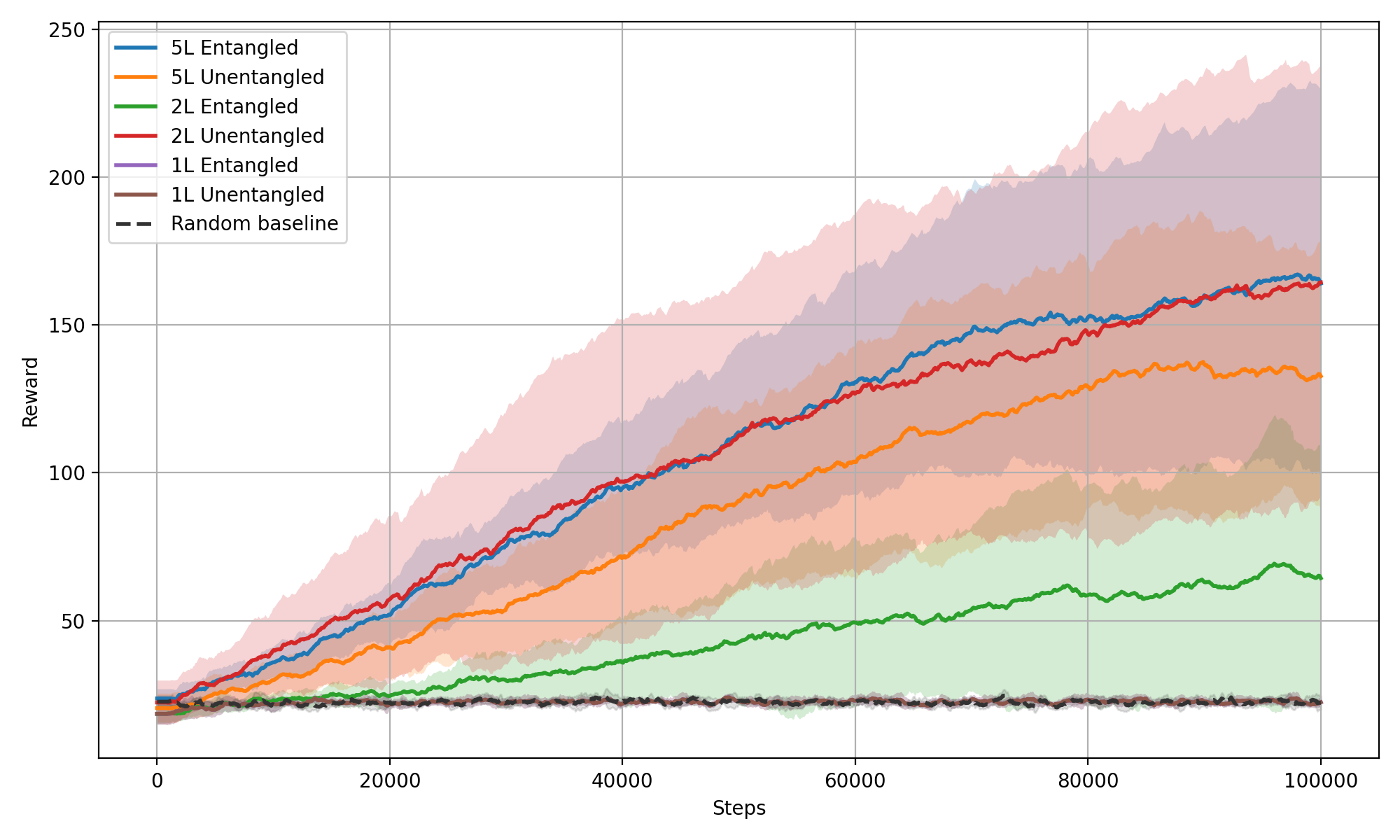}
  \caption{Effect of DR depth ($L$) on Template A with and without entanglement. Entangled circuits exhibit smoother and more consistent scaling with depth, whereas unentangled variants display irregular yet occasionally higher performance peaks, indicating poor stability.}
  \label{fig:entanglement-skolik}
\end{figure}

For Template~A, Fig.~\ref{fig:entanglement-skolik} shows the effect of applying DR to both the entangled and unentangled variants. In both cases, the expected benefits of DR are observed, namely, increased trainability and learning stability with greater circuit depth. However, important differences emerge between the two configurations. The entangled template exhibits a more monotonic progression in performance, with deeper circuits consistently achieving higher rewards. In contrast, the unentangled variant displays a more erratic pattern: although the $L=2$ configuration outperforms $L=1$ and even achieves higher mean rewards than $L=5$, it also shows greater variance across seeds. In both versions of Template~A, the single-layer case ($L=1$) fails to learn effectively, suggesting an initial BP, consistent with the trends previously observed in Fig.~\ref{fig:dr-results-skolik}.

\begin{figure*}[t]
  \centering
  % --- Subfigura (a) ---
  \begin{subfigure}[t]{0.495\textwidth}
    \centering
    \includegraphics[width=\linewidth]{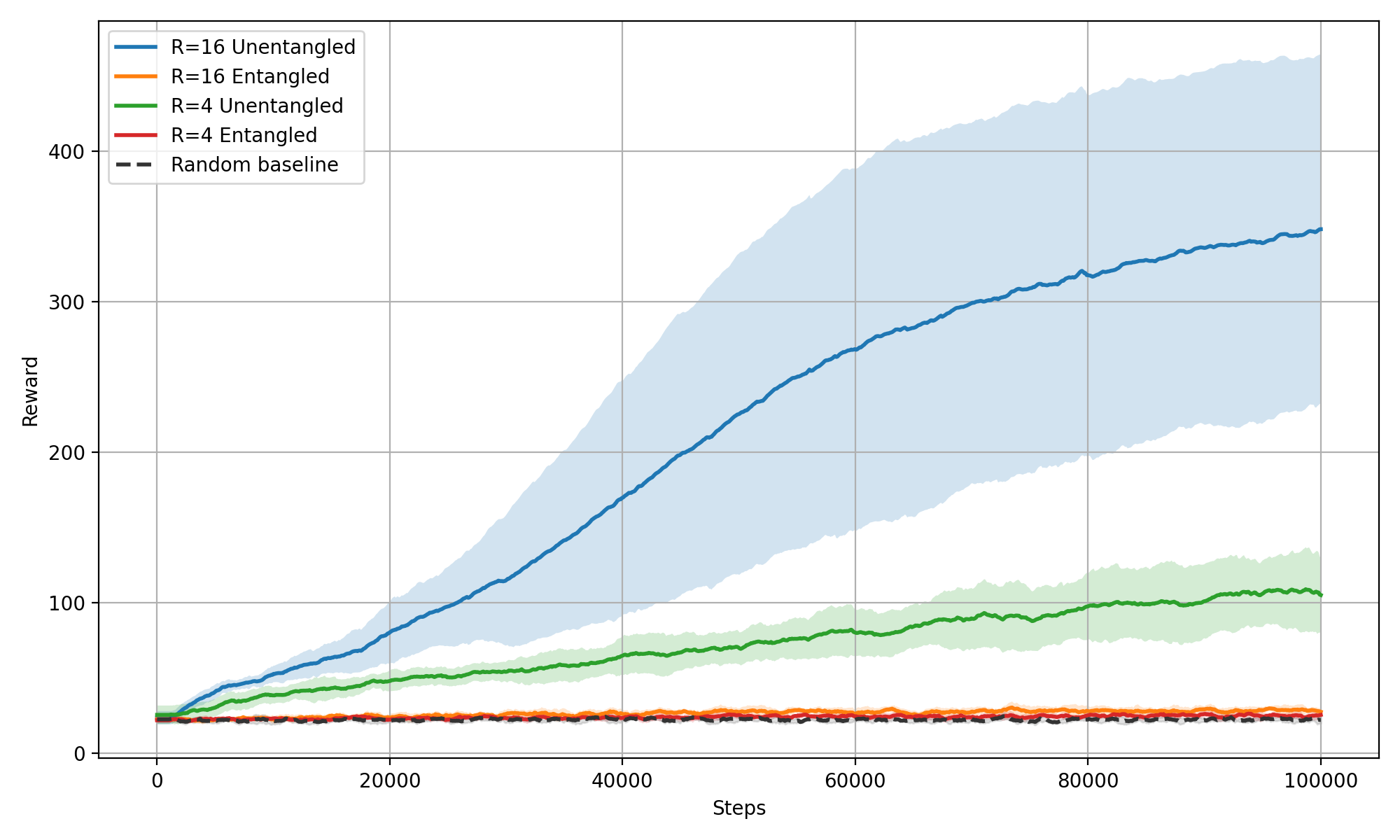}
    \caption{Effect of OR on entangled and unentangled variants of the Hsiao ansatz.}
    \label{fig:en-hsiao-or}
  \end{subfigure}
  \hfill
  % --- Subfigura (b) ---
  \begin{subfigure}[t]{0.495\textwidth}
    \centering
    \includegraphics[width=\linewidth]{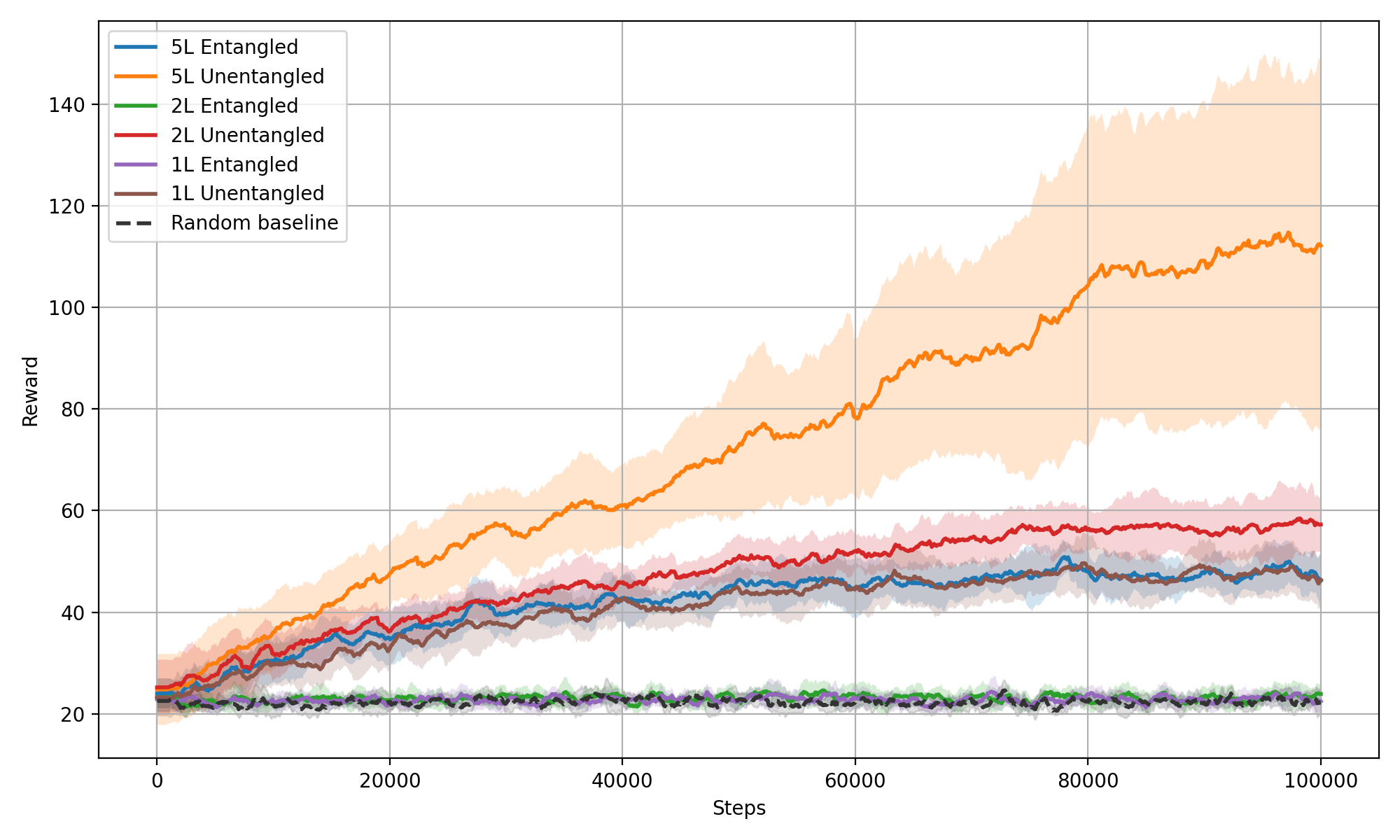}
    \caption{Effect of DR on entangled and unentangled variants of the Hsiao ansatz.}
    \label{fig:en-hsiao-dr}
  \end{subfigure}
  % --- Pie general ---
  \caption{Impact of entanglement on Template B under (a) Output Reuse and (b) Data Reuploading. Entanglement suppresses learning when combined with OR but recovers at sufficient DR depth,, highlighting architecture-specific interactions for each technique.}
  \label{fig:en-hsiao}
\end{figure*}

For Template~B, we extended the experiments to jointly examine the effect of entanglement alongside post-processing through OR and DR. Fig.~\ref{fig:en-hsiao-or} illustrates how introducing CNOT-based entanglement into an originally unentangled ansatz drastically reduces the effectiveness of OR. While the unentangled configuration behaves as expected, with noticeable gains when increasing from $R=4$ to $R=16$, the entangled counterpart fails to learn meaningfully. Entanglement is theoretically known to increase circuit expressibility by expanding the accessible subspace of the joint Hilbert space~\cite{nakaji_expressibility_2021,holmes_connecting_2022}. However, this same increase often sharpens the optimisation landscape and induces vanishing-gradient regions, manifesting as BPs during training. Our observations are consistent with this behaviour: although adding entanglement should, in principle, enrich the circuit’s representational capacity, it instead yields negligible reward progression, suggesting that the gain in expressibility is outweighed by a loss in practical trainability.

When replacing OR with DR within the same template (Fig.~\ref{fig:en-hsiao-dr}), a similar but less severe pattern emerges. Entangled variants show negligible learning for $L=1$ and $L=2$, but performance improves markedly once sufficient depth is introduced ($L=5$). By contrast, the unentangled configuration follows a predictable scaling trend consistent with prior observations, with performance improving systematically as depth increases.

The combined results highlight the significant and sometimes counterintuitive role of entanglement in the overall ansatz design for hybrid QRL. Depending on the surrounding architecture and post-processing technique, introducing or removing entanglement can yield neutral (Template~A with DR), detrimental (Template~B with OR), or consistent (Template~B with DR) effects. These outcomes indicate that structural changes in entanglement primarily modulate circuit expressibility, which, depending on depth and context, can either facilitate or obstruct trainability. Together, these findings reinforce that ansatz design choices, particularly those governing entanglement structure, critically shape trainability and its interaction with supporting techniques such as DR or OR.

\section{Discussion}
\label{sec:discussion}

This work presents, to our knowledge, the first targeted ablation of hybrid QRL pipelines that systematically isolates their three most critical components and evaluates their individual contribution. Through this decomposition, we reveal how seemingly minor implementation choices can markedly affect trainability and learning dynamics, even under a controlled and reproducible setting such as PPO–CartPole.

\paragraph{Post-PQC inference.}
Our experiments reinforce the central role of classical post-processing in hybrid pipelines. The results show that Output Reuse, although purely classical, contributes to effective learning only when coupled with a quantum block. This finding extends the work of Hsiao \emph{et~al.}~\cite{hsiao_unentangled_2022} by demonstrating that the PQC readout encodes information of sufficient structure to be exploitable by the downstream network, whereas applying OR to a classical agent yields erratic behaviour and unstable learning. In this light, OR can be regarded as a structured extension of the output-scaling techniques employed in prior work~\cite{skolik_quantum_2022,freinberger_quantum-classical_2024}, amplifying the representational capacity of the inference layer while preserving the underlying quantum correlations. These results highlight that performance gains in hybrid pipelines cannot be attributed solely to either quantum or classical components in isolation, but rather emerge from their interaction.

\paragraph{Observation embeddings.}
Our findings indicate that Data Reuploading does not perform uniformly across all angle-embedding families. The way information is injected into the PQC may act as a bottleneck, with certain embeddings providing intrinsically richer mappings. For instance, the Skolik-style embedding exhibits poor learning at shallow depths but scales consistently with both layer count and qubit width, whereas the UQC-style embedding displays higher base trainability even at minimal depth. These trends support the interpretation of DR as a mechanism that increases effective circuit capacity through layer-wise reparameterisation~\cite{coelho_vqc-based_2024}, in line with theoretical insights from single- and multidimensional Fourier analyses~\cite{schuld_effect_2021,casas_multidimensional_2023}. Furthermore, the positive scaling observed when expanding the number of qubits suggests that, when hardware permits, width augmentation can serve as an alternative to deep reuploading, mitigating barren-plateau onset through parallel representation.

\paragraph{Ansatz design and entanglement.}
The entanglement ablations confirm that ansatz architecture remains a decisive factor in QRL performance, particularly when combined with auxiliary techniques such as OR or DR. Contrary to earlier claims that ansatz design exerts only secondary influence compared with training heuristics or learning-rate schedules~\cite{kolle_study_2024}, our experiments reveal pronounced architectural dependencies. In Template~A, entanglement introduces smoother scaling with depth, whereas in Template~B, it suppresses the benefits of OR due to the likely emergence of Barren Plateaus~\cite{holmes_connecting_2022}. This heterogeneity underscores that the interplay between expressibility and trainability is highly architecture-specific. The results suggest that $SU(2)$-like constructions, though widely adopted, may conceal substantial variability in optimisation behaviour depending on the distribution of entangling gates and the dimensionality of the embedding. 

\paragraph{Broader implications.}
Overall, our findings demonstrate that trainability in hybrid QRL arises from a delicate balance between expressibility, architectural structure, and classical post-processing. Simple additive reasoning across components is insufficient: a modification that improves one block (e.g., DR depth) may be neutralised or amplified by another (e.g., entanglement or OR). Consequently, benchmarking future QRL models requires experimental designs that explicitly control for these interdependencies, ensuring that observed improvements reflect genuine quantum–classical synergy rather than artefactual pipeline interactions.

\section{Conclusions and Future Work}
\label{sec:conclusions}

This study has presented a systematic evaluation of hybrid QRL architectures by isolating the effects of post-PQC inference, observation embedding, and ansatz design under a unified PPO–CartPole framework. Each component was examined through a dedicated experimental strategy: varying the reuse factor $R$ for post-processing, increasing the reuploading depth $L$ for embeddings, and toggling entanglement within matched ansatz templates. This design allowed us to assess each block under controlled yet functionally relevant conditions, clarifying how their interaction determines overall trainability.

We have shown that post-processing techniques such as \emph{Output Reuse} (OR) enhance learning only when coupled with a meaningful quantum readout, that \emph{Data Reuploading} (DR) scales predictably but depends strongly on the embedding structure, and that entanglement can either stabilise or hinder optimisation depending on the surrounding architecture. Together, these findings provide controlled empirical evidence of how expressibility, circuit design, and classical interpretation jointly govern performance in hybrid agents.

Future work should extend this analysis to more complex environments and tasks with higher-dimensional observations, enabling assessment of scalability and robustness. While our experiments focus on gradient-based optimisation, it remains to be verified whether similar architectural dependencies persist under alternative training paradigms such as evolutionary or natural-gradient methods. Further exploration of ansatz families beyond $SU(2)$-like constructions and of hybrid architectures tailored to specific RL algorithms could yield design principles for practical QRL. Establishing statistically grounded benchmarks that account for inter-block dependencies will be essential for advancing reproducible and interpretable research in the field.

\section{Code availability}

All code used to conduct the experiments in this study is publicly available at \href{https://github.com/javier-lazaro/qrl-dissection}{github.com/javier-lazaro/qrl-dissection}. The repository accompanies this paper and includes all experiment scripts, configurations, and logged results. A self-contained copy of the \texttt{SimplyQRL} framework is provided within the repository to ensure full reproducibility.

\bibliography{bibliography}

\end{document}